\documentclass{mem}
\usepackage{natbib}\usepackage{txfonts}\usepackage{balance}
\usepackage{graphicx}
\idline{75}{282}
\begin{document}
\def\teff{$T\rm_{eff }$}
\def\kms{$\mathrm {km s}^{-1}$}

\title{
Synergy between THESEUS and E-ELT}

   \subtitle{}

\author{
E. \,Maiorano\inst{1},
L. \,Amati\inst{1},
A. \,Rossi\inst{1},
G. \,Stratta\inst{1,2,3},
E. \,Palazzi\inst{1}, 
L. \,Nicastro\inst{1} 
          }

\institute{
Istituto Nazionale di Astrofisica --
Osservatorio di Astrofisica e Scienza dello Spazio di Bologna, Area della Ricerca del CNR, Via Gobetti 101,
I-40129 Bologna, Italy
\and
Urbino University, Via S. Chiara 27, I-61029 Urbino (PU), Italy
\and
INFN-Firenze, via G. Sansone 1, I-50019 Sesto Fiorentino (FI), Italy\\
\email{maiorano@iasfbo.inaf.it}
}

\authorrunning{Maiorano}

\titlerunning{Synergy THESEUS-ELT}

\abstract{
The Transient High Energy Sky and Early Universe Surveyor (THESEUS) is a space mission concept 
aimed at exploiting Gamma-Ray Bursts for investigating the early Universe and at providing a 
substantial advancement of multi-messenger and time-domain astrophysics. 
A fundamental contribution to achieve this goal will be provided by the powerful synergy between 
THESEUS and the extremely large ground-based telescopes which will operate in the next decade, 
like E-ELT. We discuss great improvements coming from this joint effort and describe some possible 
observing scenarios.

\keywords{
Gamma-ray: bursts -- Cosmology: observations, dark ages, re-ionization, first stars -- Multimessenger:
gravitational wave, short GRBs, kilonova 
}}
\maketitle{}

\section{Introduction} 

The driving science goals of THESEUS \citep{Amati2017} aim at finding answers to multiple fundamental 
questions of modern cosmology and astrophysics, exploiting the mission unique capability to: a) explore 
the physical conditions of the Early Universe (the cosmic dawn and re-ionization era) by unveiling the 
Gamma-Ray Burst (GRB) population in the first billion years; b) perform an unprecedented deep monitoring 
of the soft X-ray transient Universe, thus providing a fundamental synergy with the next-generation of 
gravitational wave and neutrino detectors (multi-messenger astrophysics), as well as the large 
electromagnetic (EM) facilities of the next decade. Thanks to a perfect combination of its on-board 
instruments, THESEUS will be able to provide a detection of any class of GRB (long, short, X-ray flashes...), 
hugely increase the statistical sample of high-z GRBs (between 30 and 80 per year at z$>$6), detect, 
localize, and identify the electromagnetic counterparts to sources of gravitational radiation, which may be 
routinely detected in the late '20s / early '30s by next generation facilities like aLIGO/ aVirgo, KAGRA, 
Einstein Telescope and eLISA. These goals will be achieved also providing powerful synergies with the 
next-generation of extremely large multi-wavelength observatories (e.g., LSST, ELT, SKA, CTA, 
ATHENA) operating in the next decade.

In the following sections we discuss the possibility to establish a strong synergy with E-ELT, its first 
light instruments, presenting possible observational strategies and showing the reliability of such a 
cooperation and the compatibility with the queue scheduled observations of these large facilities.

\section{A basic description of E-ELT + MAORY + MICADO system} 

The European Extremely Large Telescope (E-ELT) is a revolutionary scientific project for a 40m-class 
telescope that will allow us to address many of the most pressing unsolved questions about our Universe. 
The ELT will be the largest optical/near-infrared telescope in the world and will gather 13 times more light 
than the largest optical telescopes existing today. With its innovative five-mirror design that includes 
advanced Adaptive Optics (AO), the ELT will be able to correct for the atmospheric distortions (i.e., fully 
adaptive and diffraction-limited) from the start, providing exceptional image quality, 16 times sharper 
than those from the Hubble Space Telescope.
 
Multi Conjugate Adaptive Optics RelaY (MAORY) is the adaptive optic module that will be installed at 
the E-ELT at the first light of the telescope. It provides two different types of AO correction: a very 
high correction over a small FoV (diameter $\sim$ 10 arcsec), with performances rapidly degrading 
with distance from the bright natural star used to probe the wavefront (Single Conjugate Adaptive 
Optic -- SCAO mode) and a moderate correction over a wide FoV (diameter $\sim$ 60 
arcsec), with pretty homogeneous performances over the whole FoV (Multi Conjugate 
Adaptive Optic -- MCAO mode). This will make possible to get AO assisted observations over 
a large fraction of the sky accessible from Cerro Armazones, meeting the system specification 
on Sky Coverage (SC $\ge$ 50\% over the whole sky).

While MAORY must provide a port for a second instrument, still to be defined, its main goal is 
to feed the high-resolution NIR imager and spectrograph MICADO (Multi-AO Imaging Camera for Deep 
Observations), a workhorse instrument for E-ELT.  The focal plane is equipped with 3$\times$3 detectors. 
In imaging mode MICADO will provide an option with a wide FoV (50.4 $\times$ 50.4 arcsec$^2$) with 
pixel scale of 4mas/px and a high-resolution option with a 18.9 $\times$ 18.9 arcsec$^2$ FoV and a pixel 
scale of 1.5mas/px. Long-slit spectroscopy will be available with three different slit widths: 
(a) 16 mas $\times$ 4 arcsec providing a spectral resolution R$\sim$8000 for sources 
filling the slit and 11000 $<$ R $<$ 18000, depending on wavelength, for point sources 
within the slit; 
(b) 48 mas $\times$ 4 arcsec providing R$\sim$2500 for extended 
sources filling the slit; 
(c) 20 mas $\times$ 20 arcsec with a resolution R$\sim$6000 
(when filled) specifically suited to observe galaxy nuclei, to include sky simultaneously at larger 
off- axis distances. This slit is available for the K-band only. An option on the order sorting filter 
will allow access to two different spectral ranges 0.8 --1.45 $\mu$m (IzJ passbands 
simultaneously) and 1.45--2.4 $\mu$m (HK passbands simultaneously). The pixel scale will 
always be 4 mas. 

The science cases concerning GRB science (reported in Sect. 3 and 4) are widely described in 
the MAORY Science cases White Book \citep{Fiorentino2017} together with a detailed observational 
strategy. In this paper we point out the benefit of having a strong synergy between THESEUS and E-ELT.

\section{Exploring the Early Universe with Gamma-Ray Bursts} 

Gamma-ray bursts (GRBs) are bright flashes of high-energy radiation. They are so 
luminous to be detected up to very high redshifts. By using their brightness
and their association with star forming regions within galaxies it is possible to shed 
light on the high redshift Universe. We will consider the long GRBs (LGRBs),i.e. GRBs with 
T$_{90}$ $>$ 2s associated with the death of massive stars. These major stellar explosions are 
unique powerful tracers of the star formation rate (SFR) up to z$\sim$10--12, may be 
signatures of pop-III stars. Detectable up to z $\sim$10--12 thanks to their huge X/gamma-ray 
luminosities, absorption spectroscopy of the fading NIR afterglow emission and deep imaging 
(and possible emission spectroscopy) of their host galaxies can be used to explore the re-ionization 
era, to measure the cosmic star-formation rate, the number density and properties of low-mass 
galaxies, the neutral hydrogen fraction, the escape fraction of UV photons, the cosmic chemical 
evolution.

The synergy between the extremely large ground-based telescopes available in the '20s like 
E-ELT+MAORY+MICADO and the future foreseen high-energy satellites devoted to the GRBs 
science like the THESEUS mission proposed for ESA call M5, will provide redshifts and luminosities 
that are essential to optimise the time-critical follow-up \citep{Yuan2016}. Thanks to 
this strategy, the selection of the highest priority targets will be possible for the most appropriate 
observational strategy. The present detection rate of GRBs at z $>$ 6 is about 0.7--0.8 /year, thanks 
to, and limited by, the Swift satellite capabilities and follow-up possibilities with ground and space facilities. 
This rate will likely increase to $\sim$2--4 / year at the beginning of the next decade, thanks to next 
generation space facilities like SVOM and EP (Einstein Probe), and may grow up to several tens of events 
per year at the end of the '20s if THESEUS, or an analogue mission concept, will be selected and realized 
by ESA of other space agencies.

We intend to observe both afterglows (AG) and host galaxies (HG) by performing deep imaging and 
spectroscopy with E-ELT+MAO RY+MICADO of a number of high redshift GRBs (z$>$6) triggered and 
followed-up in the first phases by THESEUS. These observations will be able to address key questions 
described in the following and can be crucial to fully characterize the properties of star-forming 
galaxies over the whole cosmic history.

\subsection{The Lyman continuum escape fraction}

\begin{figure}[]
\includegraphics[width=\columnwidth]{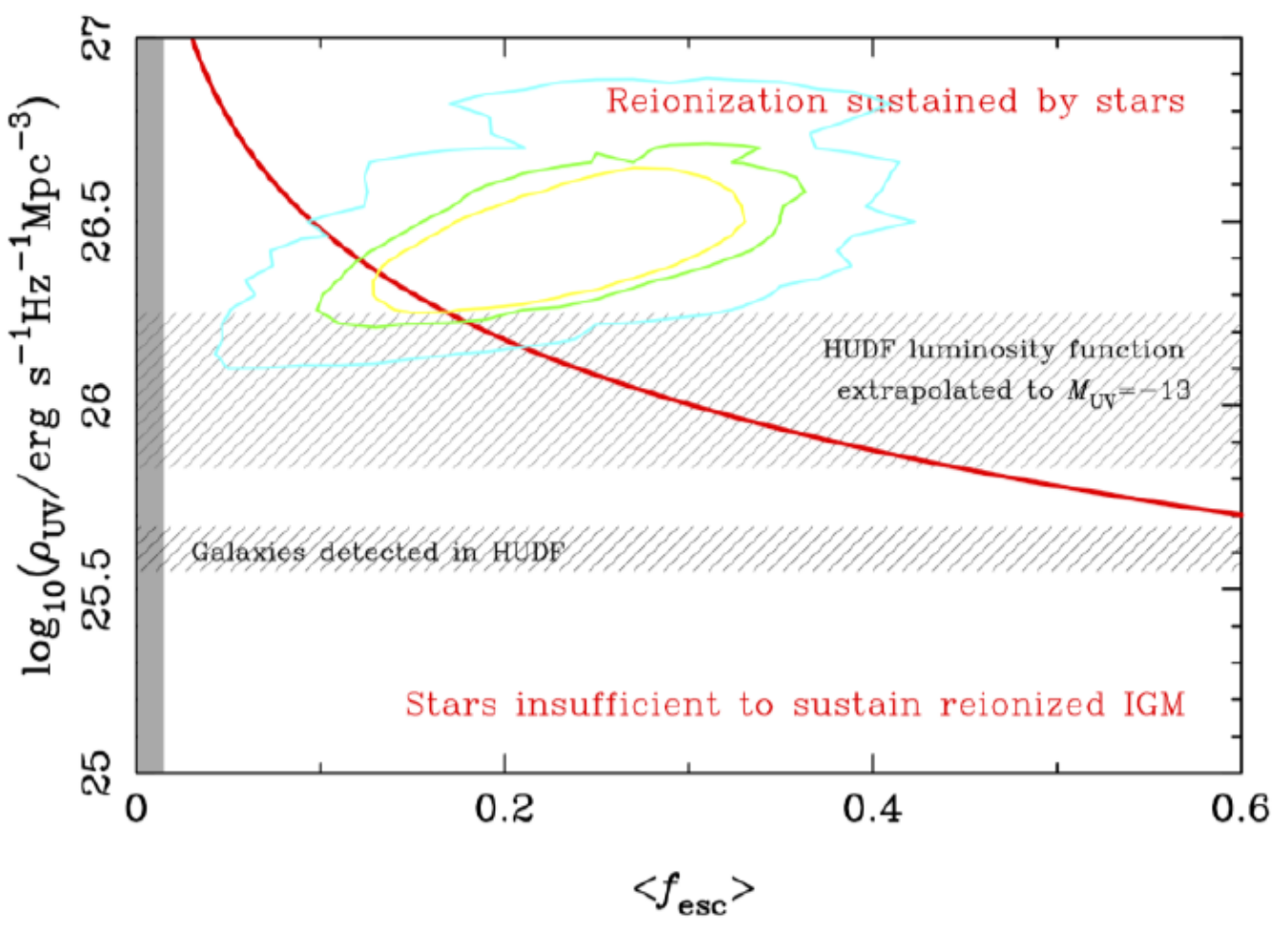}
\caption{\footnotesize
The UV luminosity density from stars at z$\sim$8 and average escape fraction $<$f$_{esc}$$>$ are insufficient 
to sustain reionization unless the galaxy luminosity function steepens to magnitudes fainter than M$_{UV}$=--13 
(grey hatched region), and/or  $<$f$_{esc}$$>$  is much higher than that typically found at z$\sim$3 (grey shaded 
region). Even in the late 2020s, $<$f$_{esc}$$>$ at these redshifts will be largely unconstrained by direct observations. 
Shown are 2-$\sigma$ contours for the cases of 10 (cyan), 25 (green) and 40 (yellow) GRBs in range z=7-9 with 
deep spectroscopic follow-up and host searches. The input parameters were log$_{10}$($\rho_{UV}$)=26.44 and  
$<$f$_{esc}$$>$ $\ge$0.23, close to the (red) borderline for maintaining reionization by stars.
}
\label{f_esc}
\end{figure}

High-S/N afterglow spectroscopy reveals the neutral hydrogen column along line-of-sight to the GRB, 
providing a powerful alternative to the lack of direct observations of escaping Lyman continuum radiation 
at z $>$ 6, especially for the small galaxies responsible for the bulk of star formation. Since the 
opacity of the medium to far ultraviolet (FUV) photons depends on this column, a statistical sample of 
afterglows can be used to infer the average escape fraction (f$_{esc}$) over many lines of sight, specifically 
to the locations of massive stars dominating global ionizing radiation production (Figure \ref{f_esc}). Useful 
constraints have so far only been possible at z = 2--4, indicating an upper limit of f$_{esc}< $7.5\% 
\citep{Fynbo2009}. Future observations of the AG of GRBs at z $>$ 6, particularly in the era of 40 m 
ground based telescopes like E-ELT+MAORY+MICADO, will provide much more precise 
constraints on the epoch of reionization.

\subsection{The build-up of metals, molecules and dust}

\begin{figure}[]
\includegraphics[width=\columnwidth]{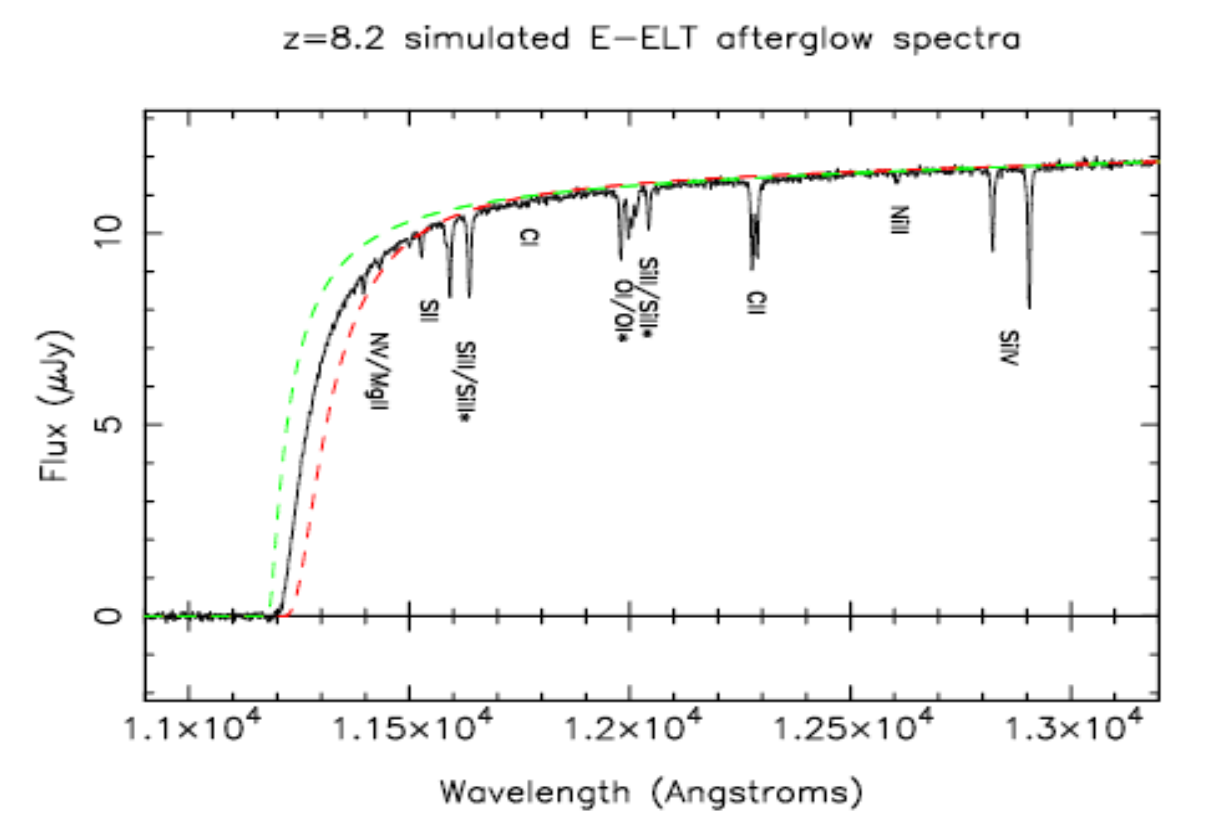}
\caption{\footnotesize
Simulated ELT 30 min spectrum of a faint GRB afterglow observed after $\sim$1 day. The S/N provides 
abundance determinations from metal absorption lines, while fitting the Lyman-$\alpha$ damping wing 
simultaneously fixes the IGM neutral fraction and the host HI column density, as illustrated by the 
two extreme models, a pure 100\% neutral IGM (green) and best-fit host absorption with a fully 
ionized IGM (red). 
}
\label{elt_spec}
\end{figure}

\begin{figure}[]
\includegraphics[width=\columnwidth]{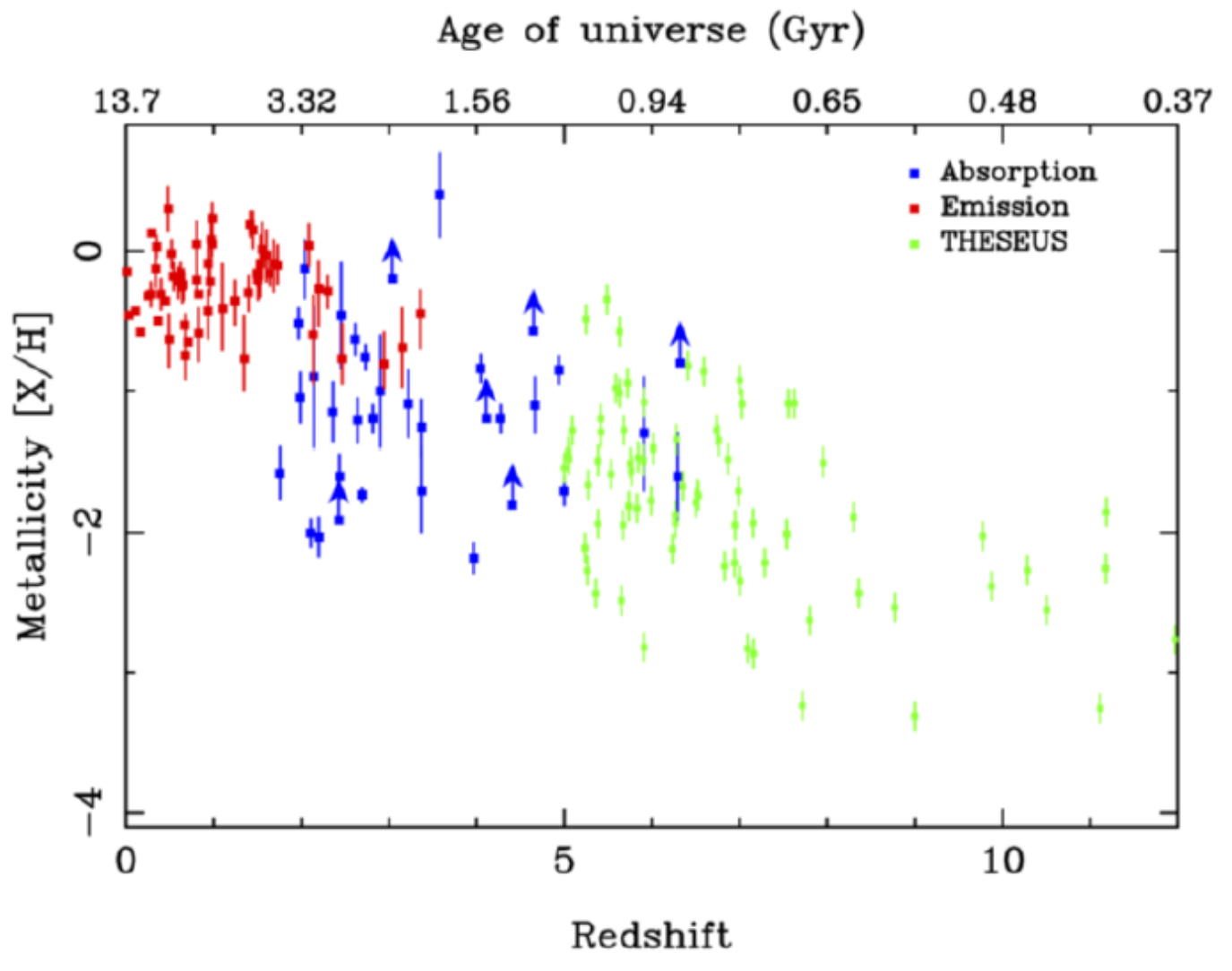}
\caption{\footnotesize
We have simulated a sample of z$>$5 THESEUS discovered pop-II GRBs with deep spectroscopic follow-up, 
and a plausible spread in metallicities and precision of determination. The metallicity distribution at z$\sim$7 
is set here to match that from the simulations of \citep{Cen2014} with a linear decline of average [X/H] with 
redshift above that. The numbers of event correspond to the minimum numbers expected from our simulations.
}
\label{metallicity}
\end{figure}

Bright GRB afterglows with their intrinsic power-law spectra provide ideal backlights for measuring not only 
the hydrogen column, but also obtaining abundances and gas kinematics probing to the hearts of their host 
galaxies \citep{Hartoog2015}. In addition, the imprint of the local dust law, and the possible detection of H$_2$ 
molecular absorption, provides further detailed evidence of the state of the host interstellar medium (ISM) 
\citep{Friis2015}. Thus they can be used to monitor cosmic metal enrichment and chemical evolution to early 
times, and search for evidence of the nucleosynthetic products of even early generations of stars (Pop-III). 
Thanks to E-ELT+MAORY+MICADO spectra, abundance determinations will be possible through simultaneous 
measurement of metal absorption lines and modelling the red-wing of Lyman-$\alpha$ to determine host HI 
column density, potentially even many days post-burst (Figure \ref{elt_spec}). As can be seen, the combination 
of the unique capabilities of THESEUS with the excellent sensitivity and spectroscopic performances of these 
future facilities would allow an unprecedented study and comparison of the average composition of the host 
galaxy and of the circumburst environment (Figure \ref{metallicity}).

\subsection{Topology of reionization}

One of the last uncharted astrophysical epochs is the time between recombination and the end of the first 
phase of star formation. Massive stars, possible progenitors of LGRBs, are likely to be a significant 
contributor to the ionizing photons at high redshift, thus GRBs are expected to occur also during the 
reionization epoch. Unlike luminous QSOs which ionize the intergalactic medium around themselves, 
GRBs do not suffer from the ``proximity effect''  \citep{Fan2006}. This makes GRBs cleaner probes of the 
Gunn-Peterson absorption used to estimate the degree of ambient ionization. With high-S/N afterglow 
spectroscopy the Lyman-$\alpha$ red damping wing can be decomposed into contributions due to the host 
galaxy and the intergalactic medium (IGM). The latter provides the hydrogen neutral fraction and so measures 
the progress of reionization local to the burst. With samples of few tens of GRBs at high redshift, we can 
begin to statistically investigate the average and variance of the reionization process as a function of redshift 
\citep{McQuinn2008}.

\subsection{Population III stars}

The multiwavelength properties of GRBs with a Pop-III progenitor are only predicted on the expected large 
masses, zero metallicity of these stars. Even the detection of a single GRB from a Pop-III progenitor would 
put fundamental constraints on the unknown properties of the first stars.

\subsection{High redshift star formation induced by galaxy interactions}

Nearby absorbers have already been identified in the spectrum of particularly bright GRB afterglows. 
They indicate possible galaxy interaction which triggers star formation (SF), and are more frequent at 
high redshift. The fraction of absorbers in the spectra of GRB afterglows is $\sim$5 times larger than in 
QSO spectra. Thus GRBs are a better tool to understand the bound between galaxy interaction and star 
formation in the high redshift Universe. We intend to exploit the high angular resolution and sensitivity 
achievable with E-ELT+MAORY+MICADO for both imaging and spectroscopy to study the morphological 
properties and the UV emission of the host of high-z GRBs triggered by THESEUS which show two strong 
nearby absorbers in the AG spectrum. These observations will allow us to test the hypothesis that galaxy 
interactions at high redshift induce the formation of very massive stars and GRB progenitors.

\subsection{HG search}

Even to the depth achieved in the Hubble Ultra-deep Field (HUDF), we only know that the faint-end of 
the luminosity function (LF) at z $>$ 6 approaches a power-law of slope $\alpha$ = 2, with an 
unconstrained cut-off at low luminosities, which affects the total luminosity integral. Although currently 
limited by small-number statistics, early application of this technique has confirmed that the majority of 
star formation at z$\sim$6 occurred in galaxies below the effective detection limit of HST 
\citep{Tanvir2012,McGuire2016}. Since the exact position and redshift of the galaxy is known via the 
GRB afterglow, GRB hosts search and observations are more efficient than equivalent deep field 
searches for Lyman-break galaxies. We thus consider fundamental to conduct deep searches with 
E-ELT+MAORY+MICADO for the hosts of GRBs at high-z in order to derive the LF at z $>$6.

\section{Characterization of EM counterpart of multi-messenger compact binary 
coalescence systems}

The launch of THESEUS will coincide with a golden era of multi-messenger astronomy. With the first 
detection of gravitational waves (GWs) by Advanced detectors \citep{Abbott2016a,Abbott2016b} a new 
window on the Universe has been opened. Several of the most powerful transient sources of GWs 
predicted by general relativity, e.g. binary neutron star (NS- NS) or NS-black hole (BH) mergers are 
expected to be associated with bright electromagnetic (EM) counterpart signals across the entire EM 
spectrum and in particular in the X-ray and gamma-ray energy bands, as well as neutrinos. They are 
thought to be the progenitors of short gamma-ray bursts (SGRBs) and kilonova/macronova transients. 
On August 17, 2017 the merger of two compact objects with masses consistent with two neutron stars 
was discovered through gravitational-wave (GW170817), gamma-ray (GRB170817A), and optical 
(SSS17a/AT 2017gfo) observations \citep{Abbott2017a,Abbott2017b,Pian2017}. In the new era of 
gravitational-wave astronomy, THESEUS aims at playing a fundamental role in multi-messenger and 
time-domain astrophysics, operating in strong synergy with future major ground-based telescopes, 
like E-ELT.

Shot GRBs (SGRBs) are identified as those GRBs with duration (T$_{90}$) less than about two seconds 
and with harder spectra with respect to the more frequent long GRBs \citep{Kouveliotou1993}. The study 
of their afterglows and host galaxies provides critical information about their explosion properties and 
progenitors. Observations carried out in the last decade show that these events, occurring in both early 
and late-type galaxies, are associated to relatively old ($\geq$ 50 Myr) stellar populations, and do not show 
evidence of associated supernovae. In addition, SGRBs have systematically larger offsets from their host 
galaxies centers than LGRBs, in line with the idea of different progenitors. Indeed, one of the key predictions 
of the compact object merger model is that systemic natal kicks may lead to substantial offsets between the 
birth and explosion sites of these systems \citep{Fryer1998,Fryer2004}. Observationally, some short GRBs 
at such large offsets can appear to be host-less because their projected locations will extend much beyond 
the visible extent of typical galaxies. The distribution of natal kick velocities of compact objects and 
compact object binaries is still an open issue, ranging from typical velocities of few hundred km/s 
\citep{Hobbs2005} to less than $\sim$30 km/s in the vast majority of the systems \citep{Beniamini2016}. 
Besides, dynamical kicks, due to close gravitational encounters, are also expected to contribute to the 
ejection of compact object binaries from their birthplace \citep{Mapelli2011}. For these reasons, it is crucial 
to assess the impact of natal kicks and dynamics on SGRBs. In particular, it is important to establish that 
the claimed host-less GRBs do not in fact lie on top of very faint system, hitherto undetected, e.g. the very 
faint host galaxy of GRB 070707 \citep{Piranomonte2008}.

Another predicted, and recently detected as said above, electromagnetic signature of NS-NS/NS-BH 
binary mergers is the so-called ``kilonova'' (KN). Hydrodynamical simulations have shown that, around 
10$^{-4}$--10$^{-2}$ M$_{\odot}$ of material may become tidally unbound and expelled \citep{Rosswog2005}. 
This material will assemble into heavy elements via r-process. Kilonova emission is thought to originate 
from the radioactive decay of these newly formed r-process elements \citep{Li1998}. Computations of the 
opacities connected to r-process material indicate that the bulk of kilonova emission is expected to peak 
in the NIR on a timescale of a few days \citep{Kasen2013,Grossman2013,Tanaka2013}. The colour and 
brightness of kilonova light curves are therefore sensitive markers of the ejecta composition, and can be 
used to gain insight into the nature and the physics of compact object mergers 
\citep{Kawaguchi2016,Kasen2015}. Kilonova emission is predicted to be isotropic and for this reason it is 
a promising electromagnetic counterpart of the merging of NS-NS/NS-BH systems, alternatively to the 
beamed SGRBs for which the chances of a joint GW-SGRB detection are rather low \citep{Ghirlanda2016}.
A kilonova counterpart is thus expected to be found among the transient sources discovered in the 
GW-localized sky regions (after the detection of a NS-NS/NS-BH GW event) or identified as an emerging 
NIR component from the rapidly decaying afterglow emission of short GRBs, as it has been done 
in the case of GRB-GW 170817 thanks to the quasi simultaneous detection of GW and SGRB.

Thanks to the synergy with THESEUS we intend to exploit the unprecedented spatial resolution and 
sensitivity of ELT+MAORY+MICADO to perform observations to shed light on the EM counterparts 
of NS-NS and NS-BH binary systems. A possible strategy to this aim is to perform deep observations to 
search and characterize the faint host galaxies of SGRBs and ``late'' ToOs (1-10 days after the burst) to
detect the kilonova emission.

\subsection{Search for kilonova emission}

\begin{figure}[]
\includegraphics[width=\columnwidth]{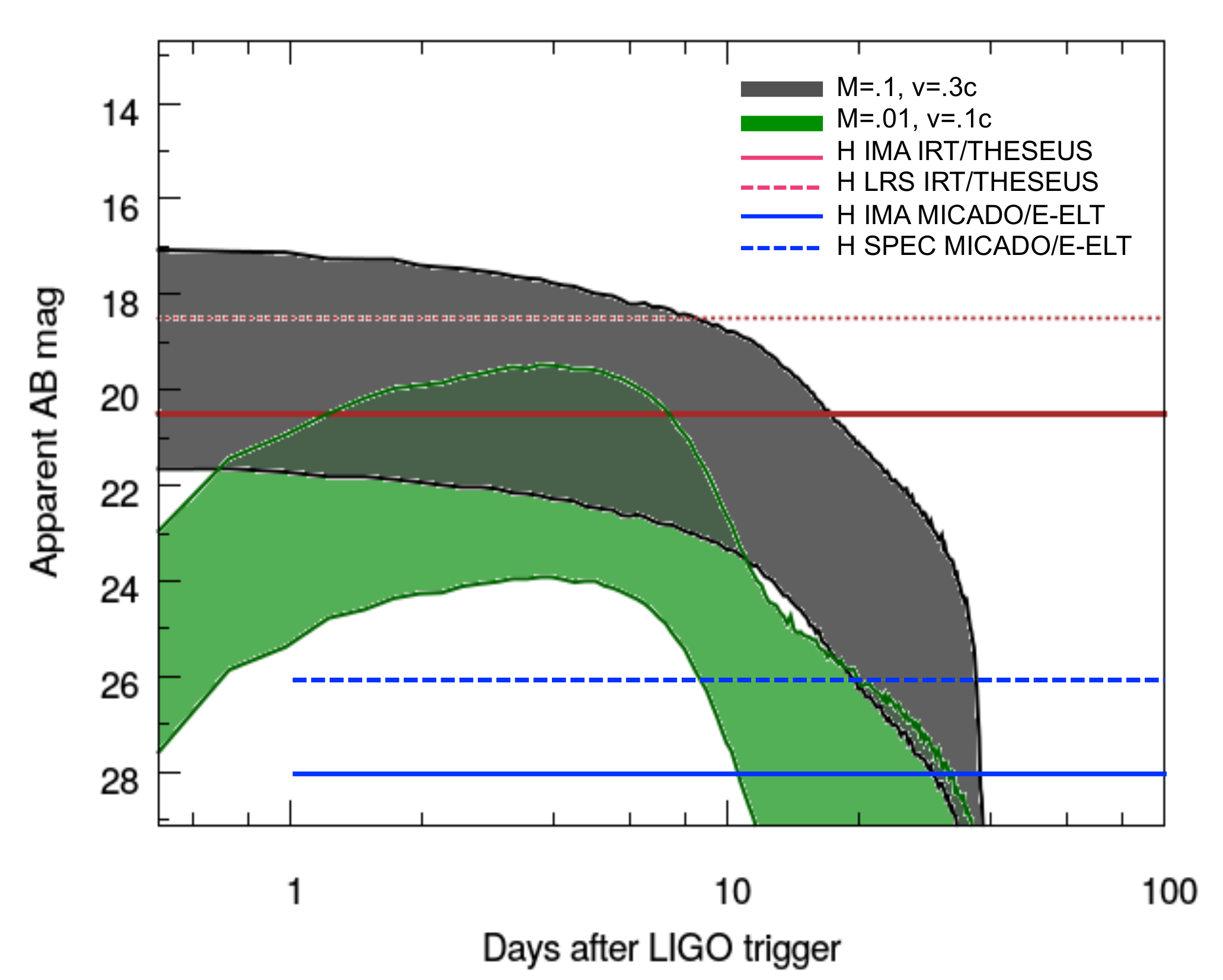}
\caption{\footnotesize
Theoretical H-band lightcurves of kilonova based on models \citep{Barnes2016}. The lightcurves 
are in observer frame for a source between 50 and 400 Mpc. Gray model is for the most optimistic case 
of a kilonova with 0.1 M$_{\odot}$ ejected mass with speed of 0.3 c. Green model is for a weaker emission, 
corresponding to 0.01 M$_{\odot}$ ejected mass with speed of 0.1 c. The continuous and dashed red lines 
indicate the THESEUS/IRT limiting H magnitudes for imaging and prism spectroscopy, respectively, with 300 
s of exposure \citep{Amati2017}. The continuous and dashed blue lines indicate the ELT/MICADO 
H-band limiting magnitudes for imaging and spectroscopy, respectively.
}
\label{KN_model}
\end{figure}

Until August 2017, the association between SGRB and kilonova emission relied only on few cases. Indeed 
the near-IR excess detected with HST observations following the SGRB 130603B \citep{Tanvir2013}
was interpreted as kilonova emission and the first direct evidence that SGRBs originate from NS-NS/NS-BH 
mergers. Late-time optical and near-IR observations place limits on the luminosity of optical and near-IR 
kilonova emission following this SGRB of $\sim$ few $\times$ 10$^{40}$ erg s$^{-1}$ \citep{Fong2016}. 
Other two suggestive pieces of evidence for SGRB/KN association has been proposed for SGRB 050709 
and SGRB 060614 \citep{Yang2015,Jin2015,Jin2016}. As discussed above, the outstanding case of 
GW170817/GRB170817A provided a fundamental confirmation of this picture. The synergy between 
THESEUS and extremely large telescopes of the future like E-ELT will allow a sistematic study and deep 
understanding of the connection and interplay between the NS-NS merging and SGRBs and KN phenomena. 
Given current kilonova models, deep optical and near-IR AG observations at 1-10 days after a SGRB to 
depths of $\sim$23-24 ABmag, up to about 200 Mpc, (i.e. the detection limit of second generation GW 
detectors) are necessary to probe a meaningful range of parameter space (Figure \ref{KN_model}). 
This highlights the key role of instruments like E-ELT+MAORY+MICADO in performing meaningful searches 
for EM counterparts to gravitational wave sources, through the dual possibility of looking for KN evidence in 
the SGRB NIR-afterglows or, alternatively, among the EM candidates found in the 1-10 deg$^{2}$ sky regions 
of GW-localized NS-NS/NS-BH mergers, that will be followed-up as soon as their position will be available 
via GCN or ATel.

\subsection{Deep observations to characterize the host galaxies of SGRBs}

NS-NS/BH mergers are expected to be found in the most massive galaxies, but the way the progenitor 
was born can also play a role. In particular, one should also consider the conditions that lead to the 
episode of star formation that generated the stellar population which the progenitor belongs to 
\citep{Leibler2010,Fong2013}. Indeed, different channels for the formation of the progenitors lead to 
different star formation histories, and thus morphology and contribution of the star clusters. Moreover, 
one of the main uncertainties on the SGRB progenitors is the time between the progenitor formation 
and the GRB explosion (delay-time). Thus, by constraining the age of the host stellar population, it is 
possible to get a reliable indication on the age of the progenitor. A first explorative study on a sample 
of SGRB host galaxies has been carried out with HST, providing constraints on their offsets and on 
how their locations track the host galaxy light distribution \citep{Fong2010}. However, up to now studies 
of the mass and age refer to the whole host galaxy, since the angular resolution (considering the 
redshift z$\sim$0.5-1) of the available facilities is not sufficient to measure age and mass of the multiple 
stellar components in the hosts. The high angular resolution and high sensitivity of MAORY+MICADO 
will allow us to resolve the morphology of the handful of SGRB hosts located at redshift z$<$0.2. Thus, 
we can study in detail any relation between the SGRB site explosion and the surrounding environments 
and estimate the mass and absolute age of star clusters that we will identify \citep{Bono2010}.

\subsection{Deep observations in the region of host-less SGRBs to search for faint host galaxies}

The short GRB offsets normalized by host-galaxy size are larger than those of long GRBs, core-collapse 
SNe, and Type Ia SNe, with only 20\% located at $\leq$1 r$_{e}$ (r$_{e}$ being the galaxy effective radius, 
which accounts for the range of HG sizes and any systematic trends in these sizes between the various 
GRB and SN populations) and about 20\% located at $\geq$5 r$_{e}$. The inferred kick velocities are 
$\sim$ 20-140 km s$^{-1}$, which is in reasonable agreement with Galactic NS-NS binaries and population 
synthesis models. About 20\% of well localized SGRBs ($\sim$7\% of the total number of SGRBs) have 
been classified as host-less \citep{Fong2013}. Several galaxies are present in the deep optical-NIR HST 
images of the region of these ``host-less'' SGRBs, but they sometimes have a high probability of chance 
coincidence because of non-optimal X-ray localization precision (2-3 arcsec). Thus, these galaxies cannot 
be identified as hosts of these events. In this context a possible strategy may first consider a small sample 
of few ($<$10) accurately localized (via radio and/or optical afterglow detection) host-less SGRBs and 
investigate if these might be inside very faint galaxies or may have been ejected by natal kicks or dynamics.
Distinguishing between these two scenarios gives constraints on the formation of NS binaries. 

\begin{figure}[]
\includegraphics[width=\columnwidth]{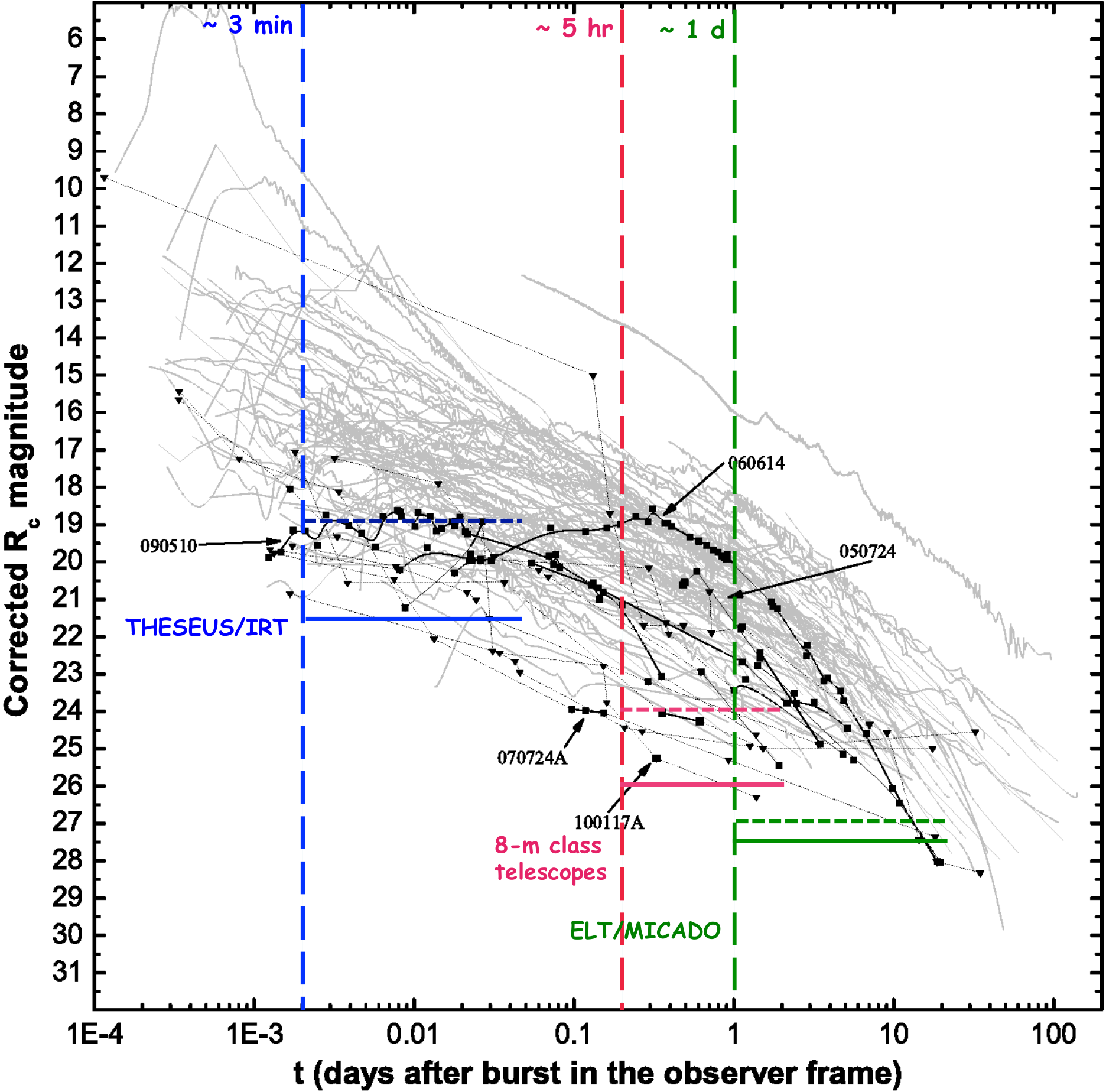}
\caption{\footnotesize
A large sample of observed R band lightcurves of long and short GRBs are shown. 
Highlighted (black lines with circles, squares and triangles) are the short GRB curves, in grey the 
long GRB ones. Data is corrected for Galactic extinction. Adapted from Kann et al. (2017). The 
limiting magnitudes achievable with THESEUS/IRT (blue lines), 8-m class telescopes (red lines) 
and ELT/MICADO (green lines) are also shown. Dashed line for the spectroscopy, solid line for 
the imaging. The magnitudes are rescaled from H-band to R-band assuming achromatic 
behaviour and a spectral index $\beta$=0.7. A tentative observation strategy could consist of the 
following steps: first starting the follow up with IRT, then activating the observations with 8-m 
class telescopes after few hours, finally, according to the brightness of the afterglow and thanks 
to the very high sensitivity of ELT, performing late observations for weeks. 
}
\label{lim_mag}
\end{figure}

\section{Possible observational strategy with E-ELT+MAORY+MICADO}

The number of GRBs at z$>$8 is expected to not exceed 10/year. Following-up 10 GRB/year with the 
major telescopes that will be available in the '30s (e.g. ELT, GMT and TMT and hopefully JWST) 
seems reliable. For 6$<$z$<$8, ``smaller'' telescopes (e.g., existing facilities like VLT) may be used, 
which will have more observing time available, thus allowing to follow tens of GRBs/year (Figure 
\ref{lim_mag}). To follow-up high-z GRBs with the major telescopes (e.g., ELT) we consider the 
Target of Opportunity (ToO) policy already widely applied at VLT or similar telescope. Due to their 
high performances, it is sufficient to ask for late ToO even or more than 1 day after the burst to have 
exquisite dataset anyway. However, note that all of the next generation ground based facilities, have 
requirements for rapid response of 30 minutes or less, similar to the rapid response mode (RRM) used 
at ESO/VLT. Alerts sent to the observatory can be acted upon rapidly via automated processes or on 
timescales as short as several minutes by the observing assistants. Some integrated fraction of time, 
possibly even as high as 10\% to 20\%, will be allotted for ToO programs but the intermissions could 
occur at anytime during queue scheduled observations. If we assume that z$\sim$8 afterglows have 
IR brightness and evolution similar to those observed up to now, the faintest afterglow will have 
H(AB)$\sim$26 at 2 hrs after the GRB (applying redshift correction to Kann et al. 2017; Figure 
\ref{lim_mag}). An afterglow with average brightness observed $\sim$1 day after the GRB will have 
similar brightness, too. Using ELT equipped with the instruments foreseen for its first light (MICADO 
and HARMONI with Adaptive Optic Correction provided by MAORY), it is possible to detect 
spectroscopically (S/N$>$3) an afterglow with H(AB)$\sim$26 with less than 1 hr exposure time. 
Similarly, considering very faint host galaxies with H(AB)$\sim$29 \citep{Savaglio2009} will be 
well detected with ELT in imaging with $\sim$1 hr exposure time. 

We report below, as examples, the set-up of observing strategy for E-ELT+MICADO+MAORY for each 
of the cases discussed above. Note that the estimates of the exposure times have been obtained with the 
ESO-ELT ETC, and all the input parameters are specified in the MAORY White Book \citep{Fiorentino2017}.

\subsection{Early Universe}

A possible strategy may be to perform the follow-up of the AGs and deep observations of the HGs for a 
number of high redshift GRBs (z$>$6). Deep imaging and spectroscopy with E-ELT+MAORY+MICADO
can allow to fully characterize the properties of star-forming galaxies over the whole cosmic history.

\subsubsection*{Afterglow follow up - request for ToO}

{\bf MICADO Pixel Scale/Fov:} Small field (1.5mas/px, 20$''$FoV) motivated by the small size of the 
actual targets. \\
{\bf MICADO Observation mode:} Standard Spectroscopy MICADO \\
{\bf Spectral set-up:} Slit 1, 16 mas $\times$ 4 arcsec, spectral coverage: IzJ+HK.\\
{\bf Estimate Number of Images/Epochs:} Since 2020 the expected rate of GRBs at z$>$6 is $>$5 per year. 
Considering visibility constraints, we expect to have $\sim$2-3 targets/yr. \\
{\bf Average Integration time per image (magnitude of targets; S/N required):} Assuming for the spectrum 
an AG at z=10, with H$\sim$26 we need $\sim$1.5hr exposure time to reach S/N$\sim$3. 
{\bf SCAO vs. MCAO:} MCAO is the only viable mode as the circumstance of a target within $\sim$10$''$ 
of a bright star (allowing SCAO) is rare or non-existing.\\
{\bf Synergies with other facilities:} Target positions will be provided by high-energy satellites available 
at that time (maybe Swift, SVOM, THESEUS) combined with accurate afterglow localizations.\\
{\bf Acquisition:} our targets are point sources so 20$''$ FoV will be selected.\\
{\bf Any other comment:} Quick response to the activations.\\

\subsubsection*{Host galaxies observations} 

{\bf MICADO Pixel Scale/Fov:} Small field (1.5mas/px, 20$''$ FoV).\\
{\bf MICADO Observation mode:} Standard Imaging and Spectroscopy\\
{\bf MICADO Spectral set-up:} slit 2, 50mas $\times$ 4arcsec, spectral coverage IzJ+HK. Slit with adjustable 
orientation would enhance the scientific return (e.g. AG+HG).\\
{\bf Filters required:} J, K plus an additional filter according to the redshift\\
{\bf Estimate Number of Images/Epochs:} At present the sample of GRB's HGs at z$>$6 is limited to 9 
events. In the late 20$'$s the expected rate of GRBs at z$>$6 is $>$5 per year. We expect to have 
$\sim$2-3 targets/yr (according to visibility constraints).\\ 
{\bf Average Integration time per image (magnitude of targets; S/N required):} Worst case for the 
imaging of an HG at K=30, with $\sim$4 hr exposure time a S/N$\sim$3 is reached. 
If the HG is brighter than K$\sim$26, a 2.5hr spectrum (low resolution) is needed to reach S/N$>$5. \\
{\bf SCAO vs. MCAO:} MCAO is the only viable mode as the circumstance of a target within $\sim$10$''$ 
of a bright star (allowing SCAO) is rare or non-existing.\\
{\bf Comparison with JWST:} Spatial resolution is the key issue here: ELT+MAORY+MICADO is much 
better than JWST by a factor of $>$3, at high SR, comparing MICADO observations in H band to JWST 
observations in I band.\\
{\bf Synergies with other facilities:} Target positions will be provided by high-energy satellites available 
at that time and/or optical telescopes (maybe Swift, SVOM, THESEUS) \\
{\bf Acquisition:} the characteristic size of GRB HG is $\sim$1 arcsec. No problem in getting a useful 
pointing within the 20$''$ FoV.\\

\subsection{Multimessenger science}

We intend to exploit the unprecedented spatial resolution and sensitivity of E-ELT +MAORY+MICADO 
to perform observations to shed light on the EM counterparts of NS-NS and NS-BH binary systems. In
the following tentative observational strategies we suggest to perform deep imaging and spectroscopy to 
search and characterize the faint host galaxies of SGRBs and "late" ToOs (1-10 days after the burst) 
to detect the kilonova emission.

\subsubsection*{Request for late ToO to search kilonova emission} 

{\bf MICADO Pixel Scale/Fov:} Small field (1.5mas/px, 20$''$ FoV) motivated by the small size of the 
actual targets.\\
{\bf MICADO Observation mode:} Standard Imaging and Spectroscopy\\
{\bf MICADO Spectral set-up:} slit 1 (16 mas $\times$ 4 arcsec), IzJ+HK Filters required: I, H. The kN 
emission is best characterized comparing lightcurves obtained with two bands separated in wavelength 
to test the color evolution; thus, we ask for I and H band (due to the better sensitivity of the instrument 
in this filter with respect to the K one).\\
{\bf Estimate Number of Images/Epochs:} uncertain rate of NS- NS, NS-BH merging systems ($\sim$few 
tens/year). Considering well-localized SGRBs at z$<$1 and/or kN candidates from GW trigger, we expect 
$\sim$3 similar targets/yr. Strategy for each event: photometric monitoring (I, H filters) of 10 epochs with 
2 days cadence starting from T0+2days and spectral acquisition for brightest events (I$<$24).\\
{\bf Average Integration time per image (magnitude of targets; S/N required):} For S/N$\sim$10, I=28, and 
H=25 are reached with integration time of 10 min and 15 s respectively. For the spectrum, with I$<$24 less 
than 2 hr exposure time are needed to reach S/N$\sim$10. \\
{\bf SCAO vs. MCAO:} the circumstances of a target within $\sim$10$''$ of a bright star (allowing SCAO) 
should be considered in order to select SCAO or MCAO mode.\\
{\bf Synergies with other facilities:} high-energy satellites available at that time (maybe Swift, SVOM, 
THESEUS), optical telescopes, ground-based GW detectors.\\

\subsubsection*{SGRBs with confirmed host galaxies} 

{\bf MICADO Pixel Scale/Fov:} Small (1.5mas/px, 20$''$ FoV) or Large (5mas/px, 50$''$x50$''$ FoV) field, 
depending on the size of the galaxy.\\
{\bf MICADO Observation mode:} Standard Imaging\\
{\bf Filters required:} J, K to characterize the mass and age of the star clusters within the host galaxy\\
{\bf Estimate Number of Images/Epochs:} 2004-2016 $\to$ 5 HGs at z$<$0.2 (about 20\% of the total SGRBs 
with measured redshift) 2 images/1 epoch per object.\\
{\bf Average Integration time per image (magnitude of targets; S/N required):} We reach K$\sim$30.5 mag 
with 2.5 hr exposure time at S/N$\sim$3 (to resolve faint star cluster inside the close SGRBs HGs) The half-light 
radii of the targets are $<$ 10 pc at z=0.2 $\to$ $\sim$point sources. \\
{\bf SCAO vs. MCAO:} MCAO is the only viable mode as the circumstance of a target within $\sim$10$''$ of a 
bright star (allowing SCAO) is unlikely.\\
{\bf Synergies with other facilities:} Targets positions and integrated magnitudes/colors from SGRBs 
detection with high-energy satellites available at that time (maybe Swift, SVOM, THESEUS) combined 
with accurate afterglow localizations.\\
{\bf Acquisition:} size of GRB HG $>$1 arcsec, $\to$ useful pointing within the 20$''$ FoV and 
50$''\times$50$''$ FoV, depending on the size of the galaxy\\
{\bf Any other comment:} possibility to observe the most peculiar targets, in spectroscopic mode with 
HARMONI/ JWST\\

\subsubsection*{Confirmed hosts-less SGRBs - search for their faint galaxies} 

{\bf MICADO Pixel Scale/Fov:} Small field (1.5mas/px, 20$''$ FoV)\\
{\bf MICADO Observation mode:} Standard Imaging\\
{\bf Filters required:} K\\
{\bf Estimate Number of Images/Epochs:} 2004-2016 $\to$ 6 host-less SGRBs and 6 ``inconclusive'' HGs 
\citep{Fong2013}. By 2024 (7 years from now) we expect a sample of$\sim$12 targets. We request 1 
images/1 epoch per object. \\
{\bf Average Integration time per image (magnitude of targets; S/N required):} Considering limiting 
magnitude for the host-less of K$\sim$25, it is possible to reach S/N$\sim$5 with 1.8 hr exposure time. 
Estimates obtained with the ESO-ELT ETC, input parameters are specified in the WB.\\
{\bf SCAO vs. MCAO:} the circumstances of a target within $\sim$1$''$ of a bright star (allowing SCAO) 
should be considered in order to select SCAO or MCAO mode.\\
{\bf Synergies with other facilities:} Targets positions from SGRBs detection with high-energy satellites 
available at that time (maybe Swift, SVOM, THESEUS) combined with accurate AG localizations. Some 
targets can be observed in spectroscopic mode with HARMONI/JWST depending on the brightness.\\
{\bf Acquisition:} the characteristic size of putative host galaxy (likely above z$\sim$1) is expected to be 
well within 20$''$ FoV. Finding charts available.\\
{\bf Any other comment:} If any host galaxy will be found, spectrum will be possibly requested.\\

\bibliographystyle{aa}

\begin{thebibliography}{}

\bibitem[Abbott et al. (2016a)] {Abbott2016a} Abbott, B.P., et al. \ 2016, PhysRevLett, 116, 241102
\bibitem[Abbott et al. (2016b)] {Abbott2016b} Abbott, B.P., et al. \ 2016, PhysRevLett, 116, 241103
\bibitem[Abbott et al. (2017a)] {Abbott2017a} Abbott, B.P., et al. \ 2017, PhysRevLett, 119, 161101
\bibitem[Abbott et al. (2017b)] {Abbott2017b} Abbott, B.P., et al. \ 2017, ApJ, 848, L13 
\bibitem[Amati et al. (2017)] {Amati2017} Amati, L., et al. \ 2017, eprint arXiv:1710.04638
\bibitem[Barnes et al. (2016)] {Barnes2016} Barnes et al. \ 2016, ApJ, 829, 110
\bibitem[Beniamini et al. (2016)] {Beniamini2016} Beniamini, P. \& Piran, T. \ 2016, MNRAS, 456, 4089
\bibitem[Bono et al. (2010)] {Bono2010} Bono et al. \ 2010, ApJ, 708, L74
\bibitem[Cen et al. (2014)] {Cen2014} Cen \& Kimm \ 2014, ApJ, 794, 50
\bibitem[Fan et al. (2006)] {Fan2006} Fan, X., et al. \ 2006, ARA\&A, 44, 415
\bibitem[Fiorentino et al. (2017)] {Fiorentino2017} Fiorentino, G., et al. \ 2017, eprint arXiv:1712.04222
\bibitem[Fong et al. (2010)] {Fong2010} Fong, Berger \& Fox, 2010, ApJ, 708, 9
\bibitem[Fong et al. (2013)] {Fong2013} Fong et al., \ 2013, ApJ, 769, 56
\bibitem[Fong et al. (2016)] {Fong2016} Fong et al. \ 2016, ApJ, 833, 151 
\bibitem[Friis et al. (2015)] {Friis2015} Friis, M., et al. \  2015, MNRAS, 451, 167 
\bibitem[Fryer et al. (1998)] {Fryer1998} Fryer, C.L., et al. \ 1998, ApJ, 496, 333
\bibitem[Fryer et al. (2004)] {Fryer2004} Fryer, C.L., et al., \ 2004, ApJ, 601, L175 
\bibitem[Fynbo et al. (2009)] {Fynbo2009} Fynbo, J.P.U., et al. \ 2009, ApJS, 185, 526
\bibitem[Ghirlanda et al. (2016)] {Ghirlanda2016} Ghirlanda, G., et al. \ 2016, A\&A, 594, A84
\bibitem[Grossman et al. (2013)] {Grossman2013} Grossman et al. \ 2013, MNRAS, 439, 757
\bibitem[Hartoog et al. (2015)] {Hartoog2015} Hartoog, O.E. et al. \ 2015, A\&A, 580, A139 
\bibitem[Hobbs et al. (2005)] {Hobbs2005} Hobbs, G., et al. \ 2005, MNRAS, 360, 974
\bibitem[Jin et al. (2015)] {Jin2015} Jin et al. \ 2015, ApJL, 811, 22
\bibitem[Jin et al. (2016)] {Jin2016} Jin et al. \ 2016, NatCo, 7, 12898
\bibitem[Kann et al. (2017)] {Kann2017} Kann, D.A. et al. \  2017, eprint arXiv:1706.00601
\bibitem[Kasen et al. (2013)] {Kasen2013} Kasen, Badnell \& Barnes \ 2013, ApJ, 774, 25
\bibitem[Kasen et al. (2015)] {Kasen2015} Kasen, Fernandez \& Metzger \ 2015, MNRAS, 450,1777
\bibitem[Kawaguchi et al. (2016)] {Kawaguchi2016} Kawaguchi et al. \ 2016, ApJ, 825, 52 
\bibitem[Kouveliotou et al. (1993)] {Kouveliotou1993} Kouveliotou, K., et al. \ 1993, ApJ, 413, L101
\bibitem[Leibler et al. (2010)] {Leibler2010} Leibler \& Berger, \ 2010, ApJ, 725, 1202
\bibitem[Mapelli et al. (2011)] {Mapelli2011} Mapelli, M., et al. \ 2011, MNRAS, 416, 1756
\bibitem[McGuire et al. (2016)] {McGuire2016} McGuire, J.T.W., et al. \ 2016, ApJ, 825, 135
\bibitem[McQuinn et al. (2008)] {McQuinn2008} McQuinn, M., et al. \ 2008, MNRAS, 388, 1101
\bibitem[Li et al.(1998)] {Li1998} Li, L. \& Paczynski, B., \ 1998, ApJ, 507, L59
\bibitem[Pian et al. (2017)] {Pian2017} Pian, E., et al. \ 2017, Nature, 551, 67 
\bibitem[Piranomonte et al. (2008)] {Piranomonte2008} Piranomonte, S., et al. \ 2008, A\&A, 491,183
\bibitem[Robertson et al. (2013)] {Robertson2013} Robertson, B.E., et al. \ 2013, ApJ, 768, 71
\bibitem[Rosswog et al. (2005)] {Rosswog2005} Rosswog, S., et al. \ 2005, ApJ, 634, 1202
\bibitem[Savaglio et al. (2009)] {Savaglio2009} Savaglio, S., et al. \ 2009, ApJ, 691, 182
\bibitem[Tanaka et al. (2013)] {Tanaka2013} Tanaka \& Hotokezaka \ 2013, ApJ, 775, 113
\bibitem[Tanvir et al. (2012)] {Tanvir2012} Tanvir, N.R., et al. \ 2012, ApJ, 754, 46
\bibitem[Tanvir et al. (2013)] {Tanvir2013} Tanvir, N.R., et al. \ 2013, Nat, 500, 547 
\bibitem[Yang et al. (2015)] {Yang2015} Yang et al. \ 2015, NatCo, 6, 7323
\bibitem[Yuan et al. (2016)] {Yuan2016} Yuan, et al. \ 2016, SSR, 202, 235



\end{thebibliography}

\end{document}